\documentclass[12pt,twoside]{article}
\usepackage{epsfig}
\usepackage{graphicx}
\usepackage{amssymb}
\title{Decomposable Medium Conditions\\ in Four-Dimensional Representation}
\author{I.V. Lindell$^1$, L. Bergamin$^2$ and A. Favaro$^3$\\
${}^1$Dept.\ Radio Science and Engng, Aalto University, Espoo, Finland,\\ ${}^2$KB\&P GmbH, Bern, Switzerland,\\ ${}^3$Dept.\ of Physics, Imperial College London, United Kingdom} 
\date{\tt ismo.lindell@tkk.fi; luzi.begamin@kbp.ch; alberto.favaro04@imperial.ac.uk}
\parskip=\medskipamount
\def\e{\begin{equation}} 
\def\f{\end{equation}} 

\def\##1{{\bf #1\mit}}
\def\%#1{{\mbox{\boldmath $#1$}}}
\def\=#1{{\overline{\overline{\mathsf #1}}}}

\def\SE{{\mathbb E}}
\def\SF{{\mathbb F}}

\def\/{\over}
\def\*{^{\displaystyle*}}

\def\.{\cdot}
\def\x{\times}
\def\:{\over}

\def\Ra{\Rightarrow}

\def\l#1{\label{eq:#1}}
\def\r#1{(\ref{eq:#1})}
\def\am{\left(\begin{array}{c}}
\def\amm{\left(\begin{array}{cc}}
\def\a{\end{array}\right)}

\def\A{\alpha}
\def\B{\beta}

\def\De{\Delta}
\def\E{\epsilon}
\def\g{\gamma}

\def\h{\eta}
\def\K{\kappa}
\def\la{\lambda}

\def\M{\mu}
\def\o{\omega}

\def\t{\tau}
\def\z{\zeta}

\def\VR{\varrho}

\def\ve{\%\varepsilon}

\def\tr{{\rm tr }}
\def\det{{\rm det}}

\def\W{\wedge}
\def\WW{\displaystyle{{}^\wedge}\llap{${}_\wedge$}}
\def\J{\rfloor}
\def\L{\lfloor}
\def\JJ{\rfloor\rfloor}
\def\LL{\lfloor\lfloor}

\begin{document}

\maketitle

\begin{abstract}
The well-known TE/TM decomposition of time-harmonic electromagnetic fields in uniaxial anisotropic media is generalized in terms of four-dimensional differential-form formalism by requiring that the field two-form satisfies an orthogonality condition with respect to two given bivectors. Conditions for the electromagnetic medium in which such a decomposition is possible are derived and found to define three subclasses of media. It is shown that the previously known classes of generalized Q-media and generalized P-media are particular cases of the proposed decomposable media (DCM) associated to a quadratic equation for the medium dyadic. As a novel solution, another class of special decomposable media (SDCM) is defined by a linear dyadic equation. The paper further discusses the properties of medium dyadics and plane-wave propagation in all the identified cases of DCM and SDCM.
\end{abstract}

\section{Introduction}

\subsection{TE/TM decomposition}

The most general linear electromagnetic medium (bi-an\-iso\-trop\-ic medium) can be expressed in terms of four medium dyadics in the three-dimensional Gibbsian vector ("engineering") representation as \cite{Kong,Methods}
\e \am \#D\\ \#B\a = \amm \=\E_g & \=\xi_g\\ \=\z_g & \=\M_g\a \.\am \#E\\ \#H\a, \l{DB}\f
where the four field vectors are elements of the vector space $\SE_1$. The number of free parameters is $4\x9=36$ in the general case. It it well known that, in a uniaxial anisotropic medium defined by medium dyadics of the form
\e \=\E_g = \E_t(\#u_x\#u_x + \#u_y\#u_y) + \E_z\#u_z\#u_z,\ \ \ \ \=\xi_g=0, \l{uniE} \f
\e \=\M_g = \M_t(\#u_x\#u_x + \#u_y\#u_y) + \M_z\#u_z\#u_z,\ \ \ \ \=\z_g=0  \l{uniM}\f
and satisfying
\e \E_t\M_z-\M_t\E_z\not=0, \l{spec}\f
any time-harmonic field with time dependence $e^{j\o t}$ can be uniquely decomposed in two parts,
\e \#E=\#E_{TE} + \#E_{TM},\ \ \ \ \#H=\#H_{TE} + \#H_{TM}, \f
satisfying 
\e \#u_z\.\#E_{TE}=0,\ \ \ \ \#u_z\.\#H_{TM}=0. \f
This property was probably first published by Clemmow in 1963 \cite{Clemmow}. In the case $\E_t\M_z-\M_t\E_z=0$ the decomposition can still be made but it is not unique. Actually, such a medium can be transformed to an isotropic medium through a suitable affine transformation \cite{Methods}. The TE/TM decomposition in isotropic media has a longer history \cite{Bromwich}.

TE/TM decomposition in the uniaxial medium can be simply demonstrated for a plane wave. In fact, because the fields of a plane wave in any medium satisfy $\#E\.\#B=0$ and $\#H\.\#D=0$, they also satisfy
\e \E_t\#E\.\#B -\M_t\#H\.\#D = (\E_t\M_z-\M_t\E_z)(\#u_z\.\#E)(\#u_z\.\#H)=0, \f
in the uniaxial medium \r{uniE}, \r{uniM}. Thus, assuming \r{spec}, the plane wave must be either a TE wave or a TM wave with respect to the axial direction defined by the unit vector $\#u_z$. Since any electromagnetic field outside the source region can be decomposed in a spectrum of plane waves, each component of which is either a TE wave or a TM wave, the field can be uniquely decomposed in TE and TM parts. The same principle is valid also for more general decompositions. Thus, it is sufficient to consider the decomposition problem for plane waves only.

\subsection{Generalized decomposition}

The TE/TM decomposition theory has been generalized to media where the fields can be decomposed to two parts satisfying either $\#a\.\#E=0$ or $\#b\.\#H=0$ where $\#a$ and $\#b$ are two given vectors \cite{Kujawski}. Even more generally, \cite{deco} analyzes the occurrence of $\#a_1\.\#E+ \#a_2\.\#H=0$ or $\#b_1\.\#E+\#b_2\.\#H=0$ where $\#a_1\cdots\#b_2$ are four given vectors. This last decomposition was shown to be possible in bi-anisotropic media with Gibbsian medium dyadics of the form 
\e \=\E_g = \frac{1}{2\t}(-\=B{}^T +\#a_2\#b_1+ \#b_2\#a_1), \l{decoE}\f
\e \=\xi_g = \#x\x\=I + \frac{1}{2\t}(\#a_2\#b_2+\#b_2\#a_2), \f
\e \=\z_g = \#z\x\=I + \frac{1}{2\h}(\#a_1\#b_1+\#b_1\#a_1), \f
\e \=\M_g = \frac{1}{2\h}(\=B +\#a_1\#b_2+ \#b_1\#a_2), \l{decoM}\f
where $\#x,\#z$ are arbitrary vectors, $\t,\h$ are arbitrary scalars and $\=B$ is an arbitrary dyadic. Media defined by \r{decoE} -- \r{decoM} have been called decomposable media \cite{deco}. The paper \cite{deco1} demonstrates that yet another scalar parameter $\A$ can be added to the definitions of the medium dyadics. In this case the definitions \r{decoE} -- \r{decoM} are generalized (after some manipulations) to the form

\e \=\E_g =  \A(\#z\x\=I + \#a_1\#b_1+\#b_1\#a_1)+ \h(-\=B{}^T +\#a_2\#b_1+ \#b_2\#a_1) , \l{decoE1} \f
\e \=\xi_g = \h(\#x\x\=I +\#a_2\#b_2+\#b_2\#a_2) + \A(\=B +\#a_1\#b_2+ \#b_1\#a_2), \f
\e \=\z_g = \t(\#z\x\=I + \#a_1\#b_1+\#b_1\#a_1)-\A (-\=B{}^T +\#a_2\#b_1+ \#b_2\#a_1), \f
\e \=\M_g = -\A (\#x\x\=I + \#a_2\#b_2+\#b_2\#a_2) +\t(\=B +\#a_1\#b_2+ \#b_1\#a_2) . \l{decoM1}\f

\subsection{Four-dimensional formalism}

Remarkably, the conditions \r{decoE1} -- \r{decoM1} of the decomposable medium can be formulated in a compact way by applying the four-dimensional differential-form formalism. The present paper thus starts directly from the four-dimensional definition of decomposable media.

It is well known that the Maxwell equations,
\e \#d\W\%\Phi =0,\ \ \ \ \#d\W\%\Psi=\%\g,  \l{Max}\f
and the medium equation
\e \%\Psi = \=M|\%\Phi, \f
have a simpler appearance in the four-dimensional differential-form representation in comparison with the three-dimensional Gibbsian vector formalism \cite{Deschamps,Hehl,Difform}. Here the electromagnetic fields are represented by two-forms $\%\Phi, \%\Psi$, elements of the space $\SF_2$, whose expansions in terms of three-dimensional two-forms $\#B,\#D$ and one-forms $\#E,\#H$ are
\e \%\Phi = \#B + \#E\W\ve_4,\ \ \ \ \ \%\Psi = \#D - \#H\W\ve_4. \f
Here, $\%\g\in\SF_3$ is the source three-form,
\e \%\g = \%\VR - \#J\W\ve_4, \f
consisting of three-dimensional charge three-form $\%\VR$ and current two-form $\#J$. In the basis of one-forms $\ve_i$, $i=1\cdots4$, $\ve_4$ denotes the temporal basis one-form. The medium dyadic $\=M\in\SF_2\SE_2$ maps two-forms to two-forms and corresponds to a $6\x6$ matrix. It is often simpler to consider the modified medium dyadic $\=M_g\in\SE_2\SE_2$ defined by
\e \=M_g = \#e_N\L\=M, \f
in terms of a quadrivector $\#e_N\in\SE_4$. The modified medium dyadic maps two-forms to bivectors, elements of the space $\SE_2$. Summary of definitions and operational rules for differential forms, multivectors and dyadics applied in this paper can be found in the appendices of \cite{JEWA02,EM06,Meta} and, more extensively, in the book \cite{Difform}.

The medium dyadic $\=M$ can be expanded in four three-dimensional dyadic components by separating terms involving the temporal vector $\#e_4$, temporal one-form $\ve_4$, or both, as
\e \=M = \=\A + \=\E'\W\#e_4 + \ve_4\W\=\M{}^{-1} + \ve_4\W\=\B\W\#e_4, \l{MM}\f
which corresponds to the matrix representation
\e \am \#D\\ \#H\a = \amm \=\A & \=\E'\\ \=\M{}^{-1} & \=\B\a|\am \#B\\ \#E\a.\l{DH} \f
One can represent the modified medium dyadic by using Gibbsian medium dyadics as
$$ \=M_g = \=\E_g\WW\#e_4\#e_4 - $$
\e -(\#e_{123}\L\=I{}_s^T+ \#e_4\W\=\xi_g)|\=\M{}_g^{-1}|(\=I\J\#e_{123} - \=\z_g\W\#e_4), \l{Mg3}\f
which corresponds to the matrix representation
\e \#e_{123}\L\am \#D\\ \#B\a = \amm \=\E_g & \=\xi_g\\ \=\z_g & \=\M_g\a|\am \#E\\ \#H\a.\l{BD} \f
The matrix is the same as in the expression \r{DB} which involves  Gibbsian vectors. 

The two sets of 3D medium dyadics have the relations \cite{Difform}
\e \=\E'= \ve_{123}\L(\=\E_g-\=\xi_g|\=\M{}_g^{-1}|\=\z_g), \l{E'}\f
\e \=\M = \ve_{123}\L\=\M_g,\ \ \ \=\M{}^{-1} = \=\M{}_g^{-1}\J\#e_{123}, \f
\e \=\A = \ve_{123}\#e_{123}\LL(\=\xi_g|\=\M{}_g^{-1}),\ \ \ \ \=\B = -\=\M{}_g^{-1}|\=\z_g. \l{AA}\f

The four-dimensional formalism allows simple definition of important classes of electromagnetic media. For example, if the modified medium dyadic can be expressed in terms of the double-wedge square of some dyadic $\=Q\in\SE_1\SE_1$ (which need not be symmetric) as
\e \=M_g = M\frac{1}{2}\=Q\WW\=Q= M\=Q{}^{(2)} , \l{MQ}\f
the corresponding three-dimensional Gibbsian dyadics satisfy relations of the form \cite{Q,Difform}
\e \=\E_g + \A\=\M{}_g^T=0,\ \ \ \=\xi{}_g^T=-\=\xi_g,\ \ \ \=\z{}_g^T=\=\z_g,  \l{Q}\f
for some scalar $\A$. Thus, $\=\E_g$ and $\=\M{}_g^T$ are multiples of the same dyadic while $\=\xi_g$ and $\=\z_g$ may be any antisymmetric dyadics. In \r{Q}, a medium defined by \r{MQ} was called a Q-medium for brevity. Such a medium is known to have the property of being non-birefringent to propagating waves. Thus, media in this class can be conceived as generalizations of isotropic media. Also, the class known as transformation media \cite{Leonhardt1,Pendry,Leonhardt2} largely coincides with the class of Q-media with a symmetric dyadic $\=Q$.

Generalizing the definition \r{Q} by adding a term
\e \=M_g = M\=Q{}^{(2)} + \#A\#B, \l{MGQ}\f
where $\#A,\#B$ are two bivectors, leads to the class of generalized Q-media, defined in \cite{genQ}. One can show that, for this kind of media, the three-dimensional medium dyadics must be of the form \r{decoE} -- \r{decoM}, i.e., any generalized Q-medium is actually a decomposable medium. However, this cannot be inverted, because the more general set of conditions \r{decoE1} -- \r{decoM1} for $\A\not=0$ cannot be achieved with medium dyadics of the form \r{MGQ}. At this stage it is not obvious how to generalize \r{MGQ} to correspond to the conditions  \r{decoE1} -- \r{decoM1}.

\subsection{Hehl-Obukhov decomposition}

In many applications a decomposition of the medium dyadic based on its symmetry properties is often useful. As shown by Hehl and Obukhov \cite{Hehl} (following the symmetry properties discussed by Post \cite{Post}), the most general medium dyadic can be uniquely decomposed in three components as \cite{Hehl},
\e \=M = \=M_1 + \=M_2 + \=M_3,  \l{M123} \f
called principal (1), skewon (2) and axion (3) parts of $\=M$. The axion part $\=M_3$ is a multiple of the unit dyadic $\=I{}^{(2)T}$ mapping any two-form to itself while the other two parts are trace free. The skewon part is defined so that the corresponding modified medium dyadic $\#e_N\L\=M_2\in\SE_2\SE_2$ is antisymmetric, while the principal part $\=M_1$ is trace free and $\#e_N\L\=M_1$ is symmetric. The decomposition \r{M123} motivates an intuitive nomenclature; for example a medium defined by $\=M=\=M_1$  is called a principal medium and one with $\=M=\=M_2+\=M_3$ is called a skewon-axion medium.

\section{Decomposable medium (DCM)}

\subsection{Plane-wave conditions}

Let us now formulate the decomposition property in terms of four-dimensional quantities. Assuming plane-wave fields
\e \%\Phi(\#x) = \%\Phi e^{\%\nu|\#x},\ \ \ \ \%\Psi(\#x) = \%\Psi e^{\%\nu|\#x}, \f
the Maxwell equations \r{Max} for the wave one-form $\%\nu$ and electromagnetic amplitude two-forms $\%\Phi, \%\Psi$ become
\e \%\nu\W\%\Phi=0,\ \ \ \ \%\nu\W\%\Psi=0. \l{nuPhi}\f
These imply the following representations in terms of potential one-forms $\%\phi,\%\psi$,
\e \%\Phi = \%\nu\W\%\phi,\ \ \ \ \%\Psi=\%\nu\W\%\psi. \l{Phinu}\f
Thus, the electromagnetic two-forms of any plane wave satisfy the orthogonality conditions
\e \%\Phi\W\%\Phi=0,\ \ \ \ \%\Phi\W\%\Psi=0,\ \ \ \ \%\Psi\W\%\Psi=0  \l{PhiPhi0}\f
in any medium. Defining the dot product between two two-forms $\%\Phi,\%\Psi$ or two bivectors $\#A,\#B$ as
\e \%\Phi\.\%\Psi=\%\Phi|(\#e_N\L\%\Psi) = \#e_N|(\%\Phi\W\%\Psi), \f
\e \#A\.\#B = \#A|(\ve_N\L\#B)=\ve_N|(\#A\W\#B), \f
the four-form conditions \r{PhiPhi0} can be expressed as the scalar conditions  
\e \%\Phi\.\%\Phi=0,\ \ \ \ \%\Phi\.\%\Psi=\%\Phi\.\=M|\%\Phi= 0,\ \ \ \ \%\Psi\.\%\Psi=\%\Phi|\=M{}^T\.\=M|\%\Phi=0. \l{Phi.Phi}\f
Thus, for any medium dyadic $\=M$, the two-form $\%\Phi$ of any plane wave satisfies a condition of the form
\e \%\Phi|(\A\#e_N\L\=I{}^{(2)T} + \B\#e_N\L\=M + \g\=M{}^T\.\=M)|\%\Phi=0,  \l{cond}\f
for arbitrary scalars $\A,\B,\g$. For the modified medium dyadic this condition becomes
\e \%\Phi|(\A\#e_N\L\=I{}^{(2)T} + \B\=M_g + \g\=M{}_g^T\.\=M_g)|\%\Phi=0,  \l{condg}\f
because we have
$$ \=M{}^T\.\=M = \=M{}^T|(\#e_N\L\=M) = (\ve_N\L\=M_g)^T|\=M_g $$
\e = \=M{}_g^T|(\ve_N\L\=M_g) = \=M{}_g^T\.\=M_g. \f

\subsection{Condition for the medium dyadic}

Let us now assume that, given two bivectors $\#A,\#B\in\SE_2$, the medium has the property that any plane wave satisfies either $\#A|\%\Phi=0$ or $\#B|\%\Phi=0$. Such waves can be respectively called A-waves and B-waves and the medium, in analogy with the medium defined by \r{decoE} --- \r{decoM}, can be called by the general name decomposable medium. Thus, any plane wave in such a medium is required to satisfy
\e( \%\Phi|\#A)(\#B|\%\Phi)=\%\Phi|(\#A\#B)|\%\Phi=0, \l{AB} \f
for two given bivectors $\#A,\#B$. Following \cite{deco} let us now define the class of decomposable media by combining \r{condg} and \r{AB} and requiring that the following condition be satisfied for all two-forms $\%\Phi$:
\e \%\Phi|(\A\#e_N\L\=I{}^{(2)T} + \B\=M_g + \g\=M{}_g^T\.\=M_g)|\%\Phi=\%\Phi|(\#A\#B)|\%\Phi.  \l{DC}\f
Since the left-hand side is zero for all media, this warrants that \r{AB} is satisfied when the medium is such that \r{DC} is satisfied. Requiring that this be valid for any two-forms $\%\Phi$ implies that the symmetric parts of the dyadics in brackets on both sides must be the same. Redefining the coefficients we can write the condition in the form
\e \A\#e_N\L\=I{}^{(2)T} + \B(\=M_g +\=M{}_g^T)+ \g\=M{}_g^T\.\=M_g=\#A\#B+\#B\#A. \l{DC1}\f
Although \r{DC1} is obviously enough to define a large class of media, it is not enough for claiming that this class covers all media for which the decomposition condition \r{AB} is satisfied. The latter question must be left for the topic of further research.

To find solutions $\=M_g$ for the condition \r{DC1}, we must separate two cases: either $\g\not=0$ or $\g=0$. 
\begin{itemize}
\item The case $\%\g\not=0$ requires solving a symmetric quadratic dyadic equation and the corresponding class of media can be called that of (proper) decomposable media or DCM for brevity. 
\item In the case $\g=0$ the quadratic dyadic equation is reduced to one of the first order. It will require separate consideration and actually defines a distinct class of media called that of special decomposable media or SDCM.
\end{itemize}

\section{Solutions to the medium conditions}

\subsection{DCM}

Assuming $\g\not=0$ in \r{DC1} we can set $\g=1$ without losing generality. In this case the DCM condition \r{DC1} can be expressed in compact form as
\e  \=M{}_g'{}^T\.\=M{}_g'  = \A'\#e_N\L\=I{}^{(2)T}, \l{gen3} \f
by defining
\e \=M{}_g'= \=M_g + \B\#e_N\L\=I{}^{(2)T} -\#D\#C, \l{Mg'} \f
\e \#C = \#A, \l{C}\f
\e \#D\. \=M_g + \B\#D- \frac{1}{2}(\#D\.\#D)\#C = \#B, \l{B}\f 
\e \A' = \B^2-\A. \l{A'}\f
The condition \r{gen3} is a quadratic dyadic equation, whose solutions are derived in the Appendix (cf. \r{M2}). Accordingly, two subclasses of DCM are obtained. The first class assumes that there exist a dyadic $\=Q\in\SE_1\SE_1$ such that $\=M{}'_g$ is a multiple of $\=Q{}^{(2)}$. The second class assumes that there exist a dyadic $\=P\in\SF_1\SE_1$ so that $\=M{}'_g$ is a multiple of $\#e_N\L\=P{}^{(2)}$. Let us consider these two cases separately and respectively call them QDCM and PDCM. This nomenclature is chosen in the light of the obvious relation with Q-media \cite{Q} and P-media \cite{P}.

Redefining $\A$, the QDCM solution of \r{gen3} as obtained from \r{Mg'} -- \r{A'} must be of the general form
\e \=M_g = \A\#e_N\L\=I{}^{(2)T}+  M\=Q{}^{(2)} + \#D\#C ,\l{QDCMg}\f
or
\e \=M = \A\=I{}^{(2)T}+ M\ve_N\L\=Q{}^{(2)} + \ve_N\L\#D\#C ,\l{DCM}\f
for some normalized dyadic $\=Q$, bivectors $\#C,\#D$ and scalars $M,\A$. Thus, the definition \r{B} is more explicitly
\e \#B =  M\=Q{}^{(2)T}\.\#D + \frac{1}{2}(\#D\.\#D)\#C.  \l{B1} \f
Given the bivectors $\#A=\#C$ and $\#B$ one can easily solve \r{B1} for $\#D$. It is also easy to verify that plane-wave fields in a medium defined by \r{DCM} satisfy the decomposition property \r{AB} (see Section IVA).

The three-dimensional Gibbsian medium dyadics corresponding to the general QDCM can be expressed in the general form \r{decoE1} -- \r{decoM1}. In fact, adding an axion term with $\A\not=0$ to the medium dyadic of the generalized Q-medium \r{MGQ} analyzed in \cite{genQ}, it can be shown to correspond to the more general class \r{QDCMg} of decomposable media \r{decoE1} -- \r{decoM1}.

The second possibility in \r{gen3}, that of PDCM, yields the following forms for the decomposable medium,
\e \=M_g = \A\#e_N\L\=I{}^{(2)T}+  M\#e_N\L\=P{}^{(2)} + \#D\#C ,\l{PDCMg}\f
or
\e \=M = \A\=I{}^{(2)T}+ M\=P{}^{(2)} + \ve_N\L\#D\#C ,\l{PDCM}\f
for some normalized dyadic $\=P$, bivectors $\#C,\#D$ and scalars $M,\A$. The definition \r{B} in this case is
\e \#B =  M\#D|\=P{}^{(2)} + \frac{1}{2}(\#D\.\#D)\#C.  \l{PB1} \f
Setting again $\#C=\#A$, one can easily solve \r{PB1} for $\#D$ in terms of given $\#A$ and $\#B$. For $\A=0$ the medium coincides with one called the generalized P-medium whose basic properties have been studied in in \cite{P}. Thus, the PDCM solution coincides with the generalized P-medium extended by an arbitrary axion component. 

Based on the expansions of the dyadic $\=P$ and the bivector product $\#D\#C$ in 3D components as
\e \=P = \=P_s + \%\pi\#e_4 + \ve_4\#p + p\ve_4\#e_4, \f 
\e \#D\#C=(\#d_1\W\#d_2+ \#d_3\W\#e_4)(\#c_1\W\#c_2+\#c_3\W\#e_4), \f
the 3D components of the medium dyadic $\=M$ expressed as \r{MM} in the PDCM take the form \cite{P}
\e \=\A = M\=P{}^{(2)}_s + \A\=I{}s^{(2)T} + (\ve_{123}\L\#d_3)(\#c_1\W\#c_2), \f
\e \=\E' = -\%\pi\W\=P_s + (\ve_{123}\L\#d_3)\#c_3, \f
\e \=\M{}^{-1} = -\=P_s\W\#p - \ve_{123}\L(\#d_1\W\#d_2)(\#c_1\W\#c_2), \f
\e \=\B = \%\pi\#p - p\=P_s -\A\=I{}_s^T -\ve_{123}\L(\#d_1\W\#d_2)\#c_3. \f
One can note that the dyadics $\=\E'$ and $\=\M{}^{-1}$ have a quite restricted form. In the case $\#D\#C=0$ they actually do not have an inverse, which is in contrast to the QDCM case. Actually, P-media and Q-media can be transformed to one another through Hodge duality \cite{P}. The same property is also valid for the generalized Q- and P-media.

\subsection{SDCM}

Let us now consider the special case of the condition \r{DC1} simplified by $\g=0$ and $\ \B=1$, which yields the following first-order dyadic equation,
\e \A\#e_N\L\=I{}^{(2)T} + \=M_g +\=M{}_g^T =\#A\#B+\#B\#A. \l{DC2}\f
Equation \r{DC2} can be interpreted so that the symmetric part of the dyadic $\=M_g -\#A\#B$ must be a multiple of $\#e_N\L\=I{}^{(2)T}$. 
Redefining $\A$, the modified medium dyadic must thus be of the form
\e \=M_g = \A\#e_N\L\=I{}^{(2)T} + \#A\#B+ \=A, \l{simple}\f
where $\=A\in\SE_2\SE_2$ is an arbitrary antisymmetric dyadic. It is known that any antisymmetric dyadic mapping two-forms to bivectors can be expressed in terms of a trace-free dyadic $\=B_o\in\SE_1\SF_1$ as \cite{IB}
\e \=A = \#e_N\L(\=I\WW\=B_o)^T,\ \ \  \ \tr\=B_o=0. \l{antisym} \f

It is now easy to verify that a medium defined by \r{simple} satisfies the decomposition condition \r{AB}. In fact, any plane wave in such a medium satisfies 
$$ \%\Phi\.\%\Psi - \A\%\Phi\.\%\Phi =  \%\Phi|(\#e_N\L\=M-\A\#e_N\L\=I{}^{(2)T})|\%\Phi $$
\e = \%\Phi|( \#A\#B+\=A)|\%\Phi = (\#A|\%\Phi)(\#B|\%\Phi)=0. \f

Alternatively,  we can replace the solution \r{simple} by
\e\=M_g = \A\#e_N\L\=I{}^{(2)T} +\=A+ \#A\#B+\#B\#A, \l{simple2}\f
which corresponds to
\e\=M = \A\=I{}^{(2)T} +\ve_N\L\=A+\ve_N\L (\#A\#B+\#B\#A). \l{simple3}\f
Without losing the generality, the bivectors can be assumed to satisfy $\#A\.\#B=0$, whence the last one of the three terms in \r{simple3} is trace free. In this case the three terms correspond to the respective axion, skewon and principal parts of the medium dyadic $\=M$. While the axion and skewon parts may be arbitrary, the principal part is restricted to be of the simple form as defined by the two bivectors $\#A$ and $\#B$. Since the principal part is not complete, i.e., it does not have an inverse, some trouble in interpreting the medium in terms of three-dimensional medium dyadics may be expected. If $\#A=\#B$ is chosen in \r{simple2} and \r{simple3}, SDCM reduces to a simplified class of media, previously called that of doubly-skew media \cite{Luzi1}.

\subsection{3D expansions for SDCM}

Because SDCM defines a novel class of decomposable media, it is interesting to find its definition in terms of three-dimensional medium parameters. Let us expand the trace-free dyadic $\=B_o$ of \r{antisym} as 
\e \=B_o = \=C_s + \#e_4\%\g_s + \#c_s\ve_4 - \#e_4\ve_4(\tr\=C_s). \f
where $\=C_s$ is a spatial dyadic, $\#c_s$ a spatial vector and $\%\g_s$ a spatial one-form. Applying
\e \=I = \=I_s + \#e_4\ve_4,\ \ \ \ \=I{}^{(2)} = \=I{}_s^{(2)} + \=I_s\WW\#e_4\ve_4, \f
where $\=I_s$ is the spatial unit dyadic, we have 
$$ \ve_N\L\=A= (\=B_o\WW\=I)^T $$
\e = (\=C_s\WW\=I_s)^T -\ve_4\W\=I{}_s^T\W\#c_s - \%\g_s\W\=I{}_s^T\W\#e_4 -\ve_4\W(\=C_s-\tr\=C_s\,\=I{}_s^T)\W\#e_4 \f
Further, we can expand
\e \#A = \#e_{123}\L\%\A_s + \#a_s\W\#e_4, \ \ \ \ \ \#B = \#e_{123}\L\%\B_s+ \#b_s\W\#e_4, \f
where the vectors $\#a_s,\#b_s$ and the one-forms $\%\A_s,\%\B_s$  are spatial. In terms of these we can write 
$$ \ve_N\L( \#A\#B+\#B\#A) = (-\ve_4\W\%\A_s + \ve_{123}\L\#a_3)(\#e_{123}\L\%\B_s+ \#b_s\W\#e_4)  $$
\e +  (-\ve_4\W\%\B_s+ \ve_{123}\L\#b_s)(\#e_{123}\L\%\A_s + \#a_s\W\#e_4). \f
Inserting the expansions in \r{simple3} and equating with \r{MM} we can identify one set of  three-dimensional medium dyadics as
\e \=\A = \A\=I{}_s^{(2)T} + (\=C_s\WW\=I_s)^T+ \ve_{123}\#e_{123}\LL(\#a_s\%\B_s+ \#b_s\%\A_s),  \l{AAA}\f
\e \=\E' = -\%\g_s\W\=I{}_s^T +\ve_{123}\L(\#a_s\#b_s+\#b_s\#a_s), \f
\e \=\M{}^{-1} = -\=I{}_s^T\W\#c_s - (\%\A_s\%\B_s +\%\B_s\%\A_s)\J\#e_{123}, \l{M-1}\f
\e \=\B = -\A\=I{}_s^T - (\=C_s-\tr\=C_s\,\=I{}_s)^T -(\%\A_s\#b_s + \%\B_s\#a_s). \l{BBB}\f
It can be seen that the dyadics $\#e_{123}\L\=\E'$ and $\=\M{}^{-1}\J\ve_{123}$ may have arbitrary antisymmetric parts while their symmetric parts are incomplete consisting of only two dyads. The decomposition \r{AB} can now be written for the 3D fields as \footnote{Here $\#B$ stands for the magnetic two-form and not for the bivector.} 
\e (\#B|(\#e_{123}\L\%\A_s) + \#E|\#a_s)(\#B|(\#e_{123}\L\%\B_s) + \#E|\#b_s) =0. \f
The 3D Gibbsian-dyadic representation corresponding to \r{AAA} -- \r{BBB} can be obtained through the transformation rules \r{E'} -- \r{AA}. It turns out that their analytic expressions become quite extensive and they are omitted here.

\subsection{Example of SDCM}

As an example, let us consider SDCM with vanishing magnetoelectric parameters:
\e \=\xi_g=\=\z_g=0,\ \ \ \ \Ra\ \ \ \ \=\A=0,\ \ \ \=\B=0. \f
This implies
\e \A=0,\ \ \ \=C_s=0,\ \ \ \#a_s\%\B_s+\#b_s\%\A_s=0, \l{ACs}\f
in the above 3D representations. Limiting to the most general case, i.e., that none of the quantities $\#a_s,\%\B_s,\#b_s,\%\A_s$ vanishes, we must have $\#b_s = \la\#a_s$ and $\%\B_s=-\la\%\A_s$ for some scalar $\la$. Thus, in this case we can write
\e \=\E' = -\%\g_s\W\=I{}_s^T + 2\la(\ve_{123}\L\#a_s)\#a_s, \f
\e \=\M{}^{-1} = -\=I{}_s^T\W\#c_s +2\la\%\A_s(\%\A_s\J\#e_{123}), \l{50} \f
and the corresponding Gibbsian dyadics become
\e \=\M_g = \frac{1}{2\la(\#c_s|\%\A_s)^2}(\#c_s\#c_s +2\la(\#c_s|\%\A_s)\#e_{123}\L(\%\A_s\W\=I{}_s^T)), \l{52}\f
\e \=\E_g =2\la\#a_s\#a_s - \#e_{123}\L( \%\g_s\W\=I{}_s^T) . \l{53}\f
The medium defined by these expressions is characterized by both electric and magnetic gyrotropy. For example, the permittivity dyadic \r{53} can be approximately realized by magnetoplasma in a high static magnetic field for low frequencies \cite{Yeh}.

\section{Dispersion equations}

To verify the decomposition of plane waves let us derive the dispersion equation for the general plane wave in all of the previous medium cases. The equation for the potential one-form can be obtained by starting from \r{nuPhi}, \r{Phinu} which yield 
\e \#e_N\L(\%\nu\W\%\Psi)= -\%\nu\J(\#e_N\L\%\Psi) = -(\%\nu\J\=M_g\L\%\nu)|\%\phi=0. \f 
This can be expressed as
\e \=D(\%\nu)|\%\phi=0,\ \ \ \ \=D(\%\nu) = \%\nu\%\nu\JJ\=M_g, \l{Dphi}\f
where $\=D(\%\nu)\in\SE_1\SE_1$ is the dispersion dyadic. The axion part of $\=M_g$ does not contribute because 
\e  \%\nu\%\nu\JJ(\#e_N\L\=I{}^{(2)T}) = \#e_N\L(\%\nu\W\%\nu\W\=I{}^T)=0, \f
and we can omit it in all medium cases. One may note that \r{Dphi} implies 
\e \=D{}^{(3)}(\%\nu)\L\%\phi= (\=D(\%\nu)|\%\phi)\W\=D{}^{(2)}(\%\nu) = 0,  \l{D3nu}\f 
which will be applied in the sequel.

\subsection{QDCM}

In the case of DCM the dispersion dyadic equals that of the generalized Q-medium,
\e \=D(\%\nu) = \%\nu\%\nu\JJ(M\=Q{}^{(2)} + \#D\#C). \f
This case has been analyzed in \cite{genQ}, but let us retrace the steps for convenience. For simplicity, let us define the vectors
\e \#d = \%\nu\J\#D,\ \ \ \ \ \#c=\%\nu\J\#C. \f
Expanding
\e \=D{}^{(3)}(\%\nu) = M^3( \%\nu\%\nu\JJ\=Q{}^{(2)})^{(3)} +M^2 ( \%\nu\%\nu\JJ\=Q{}^{(2)})^{(2)}\WW\#d\#c, \l{D3}\f
and applying the rules valid for normalized $\=Q$,
\e ( \%\nu\%\nu\JJ\=Q{}^{(2)})^{(2)} =  (\%\nu\%\nu||\=Q)\=Q{}^{(3)}= \#e_N\#e_N\LL\=Q{}^{-1T}, \f
\e ( \%\nu\%\nu\JJ\=Q{}^{(2)})^{(3)} =  (\%\nu\%\nu||\=Q)^2\=Q{}^{(4)}=  (\%\nu\%\nu||\=Q)^2\#e_N\#e_N, \f
\r{D3} can be expanded as
$$ \=D{}^{(3)}(\%\nu) = M^2 (\%\nu\%\nu||\=Q)(M (\%\nu\%\nu||\=Q)\#e_N\#e_N\LL\%\nu\%\nu + $$
$$ + (\#e_N\#e_N\LL(\=Q{}^{-1T}\WW\%\nu\%\nu)\WW\#d\#c)$$
\e = M^2(\#e_N\#e_N\LL\%\nu\%\nu ) (\%\nu\%\nu||\=Q)(M (\%\nu\%\nu||\=Q) + \=Q{}^{-1T}||\#d\#c). \f
Equation \r{D3nu} yields a scalar dispersion relation which splits in two quadratic equations as
\e \%\nu|\=Q|\%\nu=0, \l{dispA}\f
\e \%\nu|(M\=Q + \#C\#D\LL\=Q{}^{-1})|\%\nu =0.\l{dispB}\f
Actually, \r{dispA} corresponds to the A-wave and \r{dispB} to the B-wave as will be shown in the next section.

\subsection{PDCM}

Neglecting again the axion term, PDCM coincides with the generalized P-medium whose dispersion equation was derived in \cite{P}. Omitting the details, quite similar to those of the QDCM, the dispersion equation can be split in two equations which are of the form
\e \%\nu|(\#D\L\=P)|\%\nu=0,\l{nuDP}\f
\e \%\nu|(\#C\L\=P{}^{-1})|\%\nu=0. \l{nuCP}\f

\subsection{SDCM}

For SDCM the dispersion dyadic equals
\e \=D(\%\nu) = \%\nu\%\nu\JJ(\=A + \#A\#B+\#B\#A). \l{Dsimple} \f
Expressing the general antisymmetric dyadic as in \r{antisym}, where $\=B_o\in\SE_1\SF_1$ may be any trace-free dyadic, we can write
\e \%\nu\%\nu\JJ\=A= -\%\nu\J(\#e_N\L(\=B_o\WW\=I)^T\L\%\nu) = \#F\L\=I{}^T,  \l{FI}\f
where the bivector defined by
\e \#F = \#e_N\L(\%\nu\W\%\nu'),\ \ \ \ \%\nu'=\=B{}_o^T|\%\nu, \f
is simple since it satisfies $\#F\.\#F=0$. The dyadic $ \#F\L\=I{}^T\in\SE_1\SE_1$ is antisymmetric and it can be shown to satisfy 
\e (\#F\L\=I{}^T)^{(2)}=\#F\#F,\ \ \ \ (\#F\L\=I{}^T)^{(3)}=0. \f
Defining for simplicity the vectors
\e \%\nu\J\#A=\#a,\ \ \ \ \%\nu\J\#B=\#b, \l{ab1}\f
we can expand
$$ \=D{}^{(3)}(\%\nu) = (\#F\L\=I{}^T + \#a\#b+\#b\#a))^{(3)} $$
\e =(\#F\#F)\WW(\#a\#b+\#b\#a) + (\#F\L\=I{}^T)\WW(\#a\#b+\#b\#a)^{(2)}. \l{FF}\f
Applying the orthogonality $\%\nu|\#a=\%\nu|\#b=0$, we obtain
\e \#F\W\#a =  (\#e_N\L(\%\nu\W\%\nu'))\W\#a = -(\#e_N\L\%\nu)(\%\nu'|\#a), \f
and similarly for $\#F\W\#b$, whence 
\e (\#F\#F)\WW (\#a\#b+\#b\#a) = 2(\#e_N\#e_N\LL\%\nu\%\nu)(\%\nu'|\#a)(\%\nu'|\#b).\f 
The last term of \r{FF} vanishes due to
\e (\#F\L\=I{}^T)\WW (\#a\#b\WW\#b\#a) = (\#e_N\L\%\nu)\%\nu'|(\#a\#b-\#b\#a)\W(\#a\W\#b)=0. \f
Thus, we are left with 
\e \=D{}^{(3)}(\%\nu) = 2(\#e_N\#e_N\LL\%\nu\%\nu)(\%\nu'|\#a)(\%\nu'|\#b),\f 
which vanishes due to \r{D3nu}. Thus, we must have either $\%\nu'|\#a=0$ or $\%\nu'|\#b=0$ satisfied by $\%\nu'$. 

In conclusion, in the SDCM case, the fourth-order dispersion equation splits in two second-order equations 
\e \%\nu|(\=B_o\J\#A)|\%\nu=0,\ \ \ \ \ \%\nu|(\=B_o\J\#B)|\%\nu=0. \l{dispsimple} \f
The plane-wave propagation depends on the  metric dyadics $\=B_o\J\#A$ and $\=B_o\J\#B$ belonging to the space $\SE_1\SE_1$. The medium has no birefringence if the symmetric parts of these two dyadics are multiples of one another. When in addition $\#A$ is a multiple of $\#B$, the medium coincides with the doubly-skew medium of \cite{Luzi1}.

\section{Properties of the plane-wave fields}

Let us now check whether the plane-wave fields satisfy the decomposition properties associated to the corresponding media.

\subsection{QDCM}

Applying \r{QDCMg} we can write for the QDCM case
$$ \=D(\%\nu)|\%\phi = \%\nu\%\nu\JJ( M\=Q{}^{(2)} + \#D\#C - \B\#e_N\L\=I{}^{(2)T})|\%\phi$$
\e  =M(\%\nu\%\nu||\=Q)\=Q|\%\phi - M(\=Q|\%\nu)\%\nu|\=Q|\%\phi +\#d\#c|\%\phi=0. \l{phi1}\f
Since the potential one-form is not unique, one can choose an additional condition (Lorenz condition) without changing the field two-form $\%\Phi=\%\nu\W\%\phi$. Choosing the condition
\e \%\nu|\=Q|\%\phi=0, \l{Lorenz}\f
\r{phi1} is reduced to
\e  M(\%\nu\%\nu||\=Q)\=Q|\%\phi +\#d\#c|\%\phi=0. \l{phi2}\f
Now the solution corresponding to the dispersion condition \r{dispA} leads to
\e \#c|\%\phi= -\#C|\%\Phi=-\#A|\%\Phi=0, \f
because of the definition \r{C}. Thus, \r{dispA} corresponds to the A-wave.

To verify that the field corresponding to the dispersion equation \r{dispB} satisfies the B-wave condition $\#B|\%\Phi=0$ appears more complicated. Starting from $\#c|\%\phi\not=0$ and $\%\nu\%\nu||\=Q\not=0$ while assuming the condition \r{Lorenz}, from \r{phi2} we see that the potential must be of the form
\e \%\phi = \la\=Q{}^{-1}|\#d, \f
where $\la$ is some scalar coefficient. The condition \r{Lorenz} then requires that $\%\nu$ satisfy $\%\nu|\#d=0$. Invoking \r{B1} and the rule
\e 2\#D\W(\%\nu\J\#D) = \%\nu\J(\#D\W\#D) = (\#e_N\L\%\nu) (\#D\.\#D), \f
after a few algebraic steps we obtain
$$\#B|\%\Phi= \la\#B|(\%\nu\W\=Q{}^{-1}|\#d) $$
\e =   \frac{\la}{2}(\#D\.\#D)(M(\%\nu|\=Q|\%\nu) +\#c|\=Q{}^{-1}|\#d). \f
Comparing with \r{dispB}, the corresponding solution can be identified as the B-wave.

\subsection{PDCM}

Omitting again some details, in \cite{P} it has been shown that the field corresponding to the dispersion equation \r{nuCP} satisfies the condition
\e \#C|\%\Phi=0, \f
i.e., it represents an A-wave. Similarly, the solutions of the dispersion equation \r{nuDP} correspond to the field condition
\e (M\#D|\=P{}^{(2)} + \frac{1}{2}(\#D\.\#D)\#C)|\%\Phi=0, \f
which equals $\#B|\%\Phi=0$ due to the definition \r{PB1}.

\subsection{SDCM}

For the SDCM case, the equation for the potential one-form \r{Dphi} with expressions from Section IVC inserted and the axion term omitted, yields
$$ \=D(\%\nu)|\%\phi =(\#F\L\=I{}^T + \#a\#b+\#b\#a)|\%\phi $$
\e =\#e_N\L( \%\nu\W\%\nu'\W\%\phi) + \#a(\#b|\%\phi) + \#b(\#a|\%\phi)=0.  \f
Multiplying by $\%\phi|$ yields $(\#a|\%\phi)(\#b|\%\phi)=0$.  Multiplying by $\%\nu'|$ leads to the condition 
\e  (\%\nu'|\#a)(\#b|\%\phi) + (\%\nu'|\#b)(\#a|\%\phi) =0, \f
from which we obtain the relations
\e \%\nu'|\#a = \%\nu|(\=B_o\J\#A)|\%\nu=0\ \ \ \ \Ra\ \ \ \ \#a|\%\phi = -\#A|\%\Phi=0, \f
\e \%\nu'|\#b = \%\nu|(\=B_o\J\#B)|\%\nu=0\ \ \ \ \Ra\ \ \ \ \#b|\%\phi = -\#B|\%\Phi=0. \f
This shows us that the plane waves corresponding to the two dispersion equations \r{dispsimple} are respectively A-waves and B-waves.

\section{Conclusion}

In this paper we have applied four-dimensional differential-form formalism to define media in which field two-forms can be decomposed in two parts: the A-fields and the B-fields: $\%\Phi=\%\Phi_A+\%\Phi_B$. The decomposed fields are defined in terms of two given bivectors $\#A$ and $\#B$ so that they satisfy $\%\Phi_A|\#A=0$ and $\%\Phi_B|\#B=0$. Media with such a property are called decomposable media (DCM) for brevity. It is shown that a large class of four-dimensional DCM medium dyadics must satisfy an equation which is either linear or quadratic. The media satisfying the linear equation are called special decomposable media (SDCM). Those satisfying the quadratic equation are shown to fall in two subclasses which equal those previously known as generalized Q-media and generalized P-media with an added axion term. Such subclasses have been dubbed QDCM and PDCM, repectively. Basic properties of all three classes of decomposable media are discussed in the paper. The properties of QDCM and PDCM are based on earlier studies on generalized Q- and P-media while the SDCM class appears to be a novel generalization of the class of non-birefringent doubly-skew media. Dispersion equations for a plane wave are derived for all classes of media and their correspondence to the decomposition of the field are demonstrated. In the Appendix it is briefly shown that a medium dyadic satisfying the quadratic equation $\=M{}^T\.\=M= \A\#e_N\L\=I{}^{(2)T}$ must belong to either the class of Q-media or that of P-media.

\section*{Appendix: Quadratic dyadic equation}

Let us consider the following quadratic equation for the medium dyadic $\=M\in\SF_2\SE_2$,
\e \=M{}^T\.\=M = \=M{}^T|(\#e_N\L\=M) = \A\#e_N\L\=I{}^{(2)T},  \l{M2} \f
in the case $\A\not=0$. Inserting the three-dimensional expansion \r{MM}, \r{M2} equals the set of four 3D conditions
\e \=\M{}^{-1T}|(\#e_{123}\L\=\A) + \=\A{}^T |(\#e_{123}\L\=\M{}^{-1})=0, \l{1} \f
\e \=\M{}^{-1T}|(\#e_{123}\L\=\E') + \=\A{}^T |(\#e_{123}\L\=\B)= -\A\#e_{123}\L\=I{}_s^T , \l{2}\f
\e \=\B{}^T|(\#e_{123}\L\=\A) + \=\E'{}^T|(\#e_{123}\L\=\M{}^{-1})=-\A\=I_s\J\#e_{123}, \l{3}\f
\e \=\B{}^T|(\#e_{123}\L\=\E') + \=\E{}^T{}'|(\#e_{123}\L\=\B)=0. \l{4}\f
Since \r{2} equals \r{3} transposed, we can ignore \r{3}.

Now one can show that, out of the two pairs of dyadics $\=\A, \=\M{}^{-1}$ and $\=\E',\=\B$, exactly one dyadic in each pair possesses a 3D inverse. To prove this, let us first consider the pair $\=\A, \=\M{}^{-1}$ and assume that neither of the dyadics possesses an inverse. Choosing the reciprocal bases $\#e_i,\ve_i$ in a suitable manner, we can expand
\e \=\M{}^{-1}\J\ve_{123}=\ve_1\%\K_1+\ve_2\%\K_2 \in\SF_1\SF_1, \f
\e \#e_{123}\ve_{123}\LL\=\A = \#e_1\%\A_1+\#e_2\%\A_2+\#e_3\%\A_3\in\SE_1\SF_1, \f
where $\%\K_1,\%\K_2$ and $\%\A_1,\%\A_2,\%\A_3$ are one-forms of which the latter three are linearly dependent. These substituted in \r{1} yields a dyadic equation between the one-forms
\e \%\K_1\%\A_1+\%\K_2\%\A_2 = -(\%\A_1\%\K_1+\%\A_2\%\K_2). \l{K1A1}\f
Assuming that $\%\A_1\W\%\A_2\not=0$, \r{K1A1} implies that $\%\K_1$ and $\%\K_2$ occupy the same subspace as $\%\A_1$ and $\%\A_2$, and that furthermore, for some scalar $A$, we have
\e \%\K_1 = A\%\A_2,\ \ \ \ \%\K_2=-A\%\A_1. \f
The relation now becomes
\e \=\M{}^{-1} = A(\ve_1\%\A_2-\ve_2\%\A_1)\J\#e_{123}=-A\#e_3\J\=\A. \f
Substituting this in \r{2}, the resulting equation 
\e \=\A{}^T|(A(\=I{}_s^{(2)}\L\#e_3)|(\#e_{123}\L\=\E')+ \#e_{123}\L\=\B)=-\A\#e_{123}\L\=I{}_s^T \f
does not have any solutions since the left-hand side has no inverse while the right-hand side has one for $\A\not=0$. Thus, the assumption that neither $\=\A$ nor $\=\M{}^{-1}$ have an inverse is incorrect. Releasing the constraint $\%\A_1\W\%\A_2\not=0$, the case $\=\M{}^{-1}\J\ve_{123}=\ve\%\K$, $\#e_{123}\ve_{123}\LL\=\A =\#e\%\A$ can be shown to lead to the same result. Thus, at least one dyadic of the pair $\=\A,\=\M{}^{-1}$ must possess an inverse. The same conclusion is valid for the pair $\=\E',\=\B$.

To demonstrate that exactly one dyadic of each pair possesses an inverse, we notice that the left-hand sides of \r{1} and \r{4} are symmetric dyadics, whence all four terms must be antisymmetric dyadics. Because 3D antisymmetric dyadics $\=A$ can be expressed in terms of some vector $\#a$ or one-form $\%\A$ as
\e \=A\in\SE_1\SE_1,\ \ \Ra\ \ \=A = \#e_{123}\L(\%\A\W\=I{}_s^T), \f 
\e \=A\in\SE_2\SE_2,\ \ \Ra\ \ \=A = \#a\W\=I{}_s\J\#e_{123}, \f 
the dyadics in \r{1} and \r{4} must have the general form 
\e \=\A{}^T|(\#e_{123}\L\=\M{}^{-1})= \#a\W\=I{}_s\J\#e_{123}, \l{5}\f 
\e \=\B{}^T|(\#e_{123}\L\=\E') = \#e_{123}\L(\%\A\W\=I{}_s^T). \l{6}\f
Since antisymmetric 3D dyadics do not have an inverse, neither do the left-hand sides of \r{5} and \r{6}. This concludes the proof. As a conclusion, there are exactly two invertible dyadics among the four dyadics. Let us split the problem of finding solutions to \r{M2} by considering the four possible cases separately.

\subsection{Case 1: $(\=\A,\=\B)$}

Assuming that $\=\A$ and $\=\B$ possess inverses, the other two dyadics can be solved from \r{5} and \r{6} as
\e  \=\M{}^{-1} = (\ve_{123}\#e_{123}\LL\=\A{}^{-1T})\W\#a,  \l{M}\f
\e \=\E' = (\ve_{123}\#e_{123}\LL\=\B{}^{-1T})\L\%\A. \l{E}\f
Defining the dyadic
\e \=X =  (\ve_{123}\#e_{123}\LL\=\A{}^T)|\=\B\ \ \ \in\SF_1\SE_1, \l{X1}\f
the inverse of which exists due to the above assumption, we can express \r{2} as a condition for the dyadic $\=X$,
\e \ve_{123}\#e_{123}\LL(\=X{}^{-1T}\WW\#a\%\A) + \=X = -\A\=I{}_s^T. \l{XX1}\f
Now it is easy to check that the solution of \r{XX1} must be of the uniaxial form
\e \=X = A\=I{}_s^T + \frac{1}{A}\%\A\#a,\ \ \ \Ra\ \ \ \det\=X= A(A^2 + \#a|\%\A), \l{X} \f
where the parameter $A$ and the scalar $\A$ of \r{M2} are related as
\e \A = -\frac{1}{A}(A^2+\#a|\%\A)= \frac{1}{A^2}\det\=X. \l{alpha}\f
At this point we can construct the medium dyadic $\=M$ satisfying \r{M2} for Case 1 from \r{X1}, \r{M} and \r{E} as
\e \=M = \frac{1}{A^2\det\=\B} (\det\=X\=\B{}^{(2)}
 - A\=\B\WW(\=\B|\%\A + \det\=\B\, \ve_4)(\#a +A\#e_4)), \l{Mcase1}\f
and it can be further expressed in the compact form
\e \=M = M \=P{}^{(2)},\ \ \ \ M = \frac{1}{A^2\det\=X\det\=\B}, \l{MP}\f
\e \=P = (\det\=X)\=\B - A(\=\B|\%\A + \det\=\B\, \ve_4)(\#a +A\#e_4). \l{P}\f
Media defined by medium dyadics of the form $M\=P{}^{(2)}$ for some dyadic $\=P$ have been called P-media \cite{P}.

\subsection{Case 2: $(\=\E',\=\M{}^{-1})$}

Assuming that $\=\E'$ and $\=\M{}^{-1}$ possess inverses, the other two medium dyadics can be expressed as
\e \=\A = -(\ve_{123}\#e_{123}\LL\=\M{}^T)\W\#a, \l{A1}\f 
\e \=\B = -(\ve_{123}\#e_{123}\LL\=\E'{}^{-1T})\L\%\A. \l{B2} \f
Defining the dyadic
\e \=X = (\ve_{123}\#e_{123}\LL\=\M{}^{-1T})|\=\E' \ \in\SF_1\SE_1, \l{X2}\f
\r{2} can be shown to yield \r{XX1} in spite of the different definition of $\=X$. Thus, the expression \r{X} can be applied. Let us for simplicity define the additional dyadic
\e \=Y = \#e_{123}\L\=\M\ \in\SE_1\SE_1,\ \ \ \ \De_Y=\ve_{123}\ve_{123}||\=Y^{(3)} \f
in terms of which we can express from \r{X2}, \r{A1} and \r{B2} the dyadics
\e \=\E'= \ve_{123}\L\=Y{}^T|\=X, \f
\e \=\A = -\ve_{123}\L\=Y{}^T\W\#a, \f 
\e \=\B =  -\frac{1}{\De_Y\det\=X}\ve_{123}\L\=Y{}^{(2)T}|\=X{}^{(2)}\L\%\A.  \f

Substituting these in \r{MM}, the modified medium dyadic $\=M_g=\#e_N\L\=M$ becomes
\e \=M_g = -\frac{1}{A^2\De_Y}(\=Y{}^T|\=X)^{(2)}  +  
(\=Y{}^T|\=X)\WW(\frac{\=Y{}^T|\%\A}{A^2\De_Y}-\#e_4)(\frac{\#a}{A}-\#e_4).  \l{case2b}\f
This can be expressed in the compact form
\e \=M_g = M\=Q{}^{(2)},\ \ \ \ M =  -\frac{1}{A^2\De_Y},  \l{MQ2}\f
\e \=Q =\=Y{}^T|\=X - \frac{1}{A}(\=Y{}^T|\%\A-A^2\De_Y\#e_4)(\#a- A\#e_4). \f
Media defined by modified medium dyadics of the form $\=M_g=M\=Q{}^{(2)}$ for some dyadic $\=Q$ have been called Q-media \cite{Q}.

\subsection{Case 3: $(\=\A,\=\E')$}

As a third case we consider the possibility that $\=\A$ and $\=\E'$ have inverses. Equation \r{2} now becomes
\e \#a\W(\#e_{123}\L\=\A{}^{-1}|\=\E') + (\=\A{}^T |\=\E'{}^{-1T}\J\#e_{123})\L\%\A= \A\#e_{123}\L\=I{}_s^T . \l{case3}\f
Denoting this time
\e \=X = \#e_{123}\L\=\A{}^{-1}|\=\E' \in\SE_1\SE_1,\ \ \ \ \De_X = \ve_N\ve_N||\=X{}^{(3)}, \f
which has an inverse within the limits of Case 3, the condition \r{case3} takes the form
\e \#a\W\=X + (\#e_{123}\#e_{123}\LL\=X{}^{-1T})\L\%\A= \A\#e_{123}\L\=I{}_s^T. \l{aX}\f
Multiplying by $|\%\A$ we obtain
\e \#a\W\=X|\%\A = \A\#e_{123}\L\%\A,\ \ \ \ \Ra\ \ \ \#a|\%\A=0, \f
whence \r{aX} becomes
\e \#b\W\=X  = \A\#e_{123}\L\=I{}_s^T,\ \ \ \ \#b = \#a + \frac{\=X|\%\A}{\De_X}.\f
Because of the assumption $\A\not=0$, this condition leads to an impasse: the dyadic on the right-hand side has an inverse while that on the left-hand side has not. Thus, the original assumption that $\=\A$ and $\=\E'$ possess inverses is obviously invalid.

\subsection{Case 4: $(\=\B,\=\M{}^{-1})$}

As the final case we consider the possibility that $\=\B$ and $\=\M{}^{-1}$ have inverses. Equation \r{2} now becomes
\e (\=\M{}^{-1T}|\=\B{}^{-1T}\J\#e_{123})\L\%\A + \#a\W(\#e_{123}\L\=\M|\=\B)= -\A\#e_{123}\L\=I{}_s^T. \l{case4}\f
Since this is of the same form as \r{case3}, the conclusion must be similar: the assumption that $\=\B$ and $\=\M{}^{-1}$ have inverses cannot be valid for $\=M$ to satisfy \r{M2} with $\A\not=0$.

In conclusion, there are only two classes of solutions to the quadratic equation \r{M2}, that of Q-media and that of P-media.

\end{document}